\RequirePackage{amsmath}
\documentclass[conference]{IEEEtran}
\usepackage{graphicx}
\usepackage{tikz} 
\usetikzlibrary{positioning}
\usetikzlibrary{arrows,matrix}
\usetikzlibrary{trees,shapes,decorations}  
\usepackage{orcidlink}
\usepackage{paralist} 
\usepackage{xcolor}
\usepackage{url}
\usepackage{subfig}
\usepackage{pgf-pie}
\usepackage{comment}

\definecolor{red}  {rgb}{0.9,0.0,0.0}
\definecolor{green}{rgb}{0.0,0.7,0.0}
\definecolor{blue} {rgb}{0.0,0.0,0.9}

\usepackage{amsfonts}
\usepackage{amsmath,mathtools,commath,nicefrac}

\usepackage{graphicx,float}

\usepackage{tabularx}
\usepackage{booktabs}

\usepackage{amsthm}

\usepackage{tikz}
\usepackage{comment}
\newtheorem{theorem}{Theorem}
\newtheorem{corollary}{Corollary}
\newtheorem{definition}{Definition}
\newtheorem{cnstr}{\textbf{Construction}}

\theoremstyle{definition} 

\newcommand{\extended}[1]{}

\usepackage[obeyclassoptions,mode=buildnew]{standalone}
\begin{document}
\title{Majority is not Needed: \\ A Counterstrategy to Selfish Mining}

	\author{
\IEEEauthorblockN{
Jonathan Gal
}
\IEEEauthorblockA{
Technion
}

\and

\IEEEauthorblockN{
Maytal Bracha Szabo
}
\IEEEauthorblockA{
Technion
}

\and

\IEEEauthorblockN{
Ori Rottenstreich
}
\IEEEauthorblockA{
Technion
}
}

         \IEEEoverridecommandlockouts
\IEEEpubid{\makebox[\columnwidth]{979-8-3315-4135-4/25/\$31.00~\copyright2025 IEEE \hfill} \hspace{\columnsep}\makebox[\columnwidth]{ }}
	\maketitle	
    \IEEEpubidadjcol\begin{abstract}
    In recent years, several works have investigated selfish mining attacks, where a miner or group of miners withhold blocks to gain an advantage over honest miners.
        Most of these studies assumed that miners outside the selfish mining pool would continue to mine honestly.
	    However, remaining honest is typically not incentive-compatible, especially if another pool is using selfish mining or a similar strategy. In this paper, we reveal the opportunity for a sufficiently large pool to exploit another selfish mining pool, enabling it to monopolize profits and launch 51\% attacks with less than half the total computing power. We show that beyond selfish mining, this strategy can leverage any approach that deviates from honest mining.
\end{abstract}
	
	\section{Introduction}
	Bitcoin, a peer-to-peer electronic cash system, was designed 
    to enable direct payment between semi-anonymous clients without a central financial institution~\cite{NakamotoBitcoin}. This is achieved by a distributed blockchain that relies on a community of 
	\textit{miners}, who often form mining pools to combine their computational power and share rewards. These pools 
	maintain the chain by providing proof-of-work (puzzle solving) to record transactions. A miner has a {\em mining power}, a limited amount of computational power.
    Miners are rewarded with transaction fees and newly minted bitcoins, with the bitcoin algorithm dynamically adjusting mining difficulty to stabilize the total profit, regardless of total computing power.
    
	To maintain the blockchain's integrity, Bitcoin resolves occasional forks—caused by simultaneous puzzle-solving—by adopting the longest branch as the main chain while pruning others. Miners’ revenue depends on the fraction of blocks they contribute to the surviving chain, making mining efficiency crucial. Intuitively, miners seem incentivized to work solely on the longest branch to maximize their rewards.
    However, Eyal and Sirer \cite{main} disprove this notion, showing that a pool of miners can increase their relative income by adopting a malicious strategy known as \emph{Selfish Mining}, a strategy where the selfish miners artificially create forks to reduce the number of block accepted into the blockchain by the other miners.
	
	In this work we propose a new strategy, called the {\em piggyback strategy}, which takes advantage of a form of mining we denote as {\em deviant mining}. This strategy collects rewards by utilizing a side-effect of selfish mining to gain complete control over the blockchain. We offer the following three concrete contributions: 
	\begin{itemize}
	    \item 	The ``piggyback" strategy - a mining strategy  to achieve 51\% capabilities with less then half of the computing power
	    \item Determining the optimal wait times for piggybackers to reveal their branch
	    \item Analyzing available responses to deviant mining pools
	\end{itemize}
	\section{Background and Related Work}
	\label{sec:related}
    Several recent works analyze strategies that exploit the blockchain protocol to gain a bigger fraction of the revenue than their computing power. 
       A major exploit is selfish mining~\cite{main}, aimed to cause honest miners (miners who are not part of the pool) to perform wasted computations on what they see as the longest chain. When the pool reveals its longer branch it invalidates the honest miner's blocks, wasting their resources and gaining a greater fraction of valid blocks (and revenue) than their mining power.
       
	 Detecting selfish mining was also investigated~\cite{detect, detection, statistical}. When attacks of this type occur the amount of blocks that are mined and then discarded sharply increases. When all miners are presumably honest the amount of discarded blocks is low (about 1\% for Bitcoin~\cite{stale}).
     
    Selfish mining becomes profitable only after several weeks of operation~\cite{main}, providing other participants enough time to detect and potentially counter the attack before it becomes financially viable for the malicious miner.
    
	An extension explored in the literature~\cite{multiC,multiUK} involves multiple pools attempting selfish mining simultaneously. It was shown that the safety level, the minimum fraction of power required by honest miners to prevent selfish mining, remains the same regardless of the number of selfish pools~\cite{multiUK}, allowing us to treat them as a single pool.
	
	The double-spending attack~\cite{doubSpnd} involves executing a transaction on a private branch where the transaction is omitted. The attacker races against the public chain, and once the private branch surpasses or matches the public one, the attacker publishes it, invalidating the original transaction. When the attacker controls more than half of the total computing power, they can consistently perform double-spending, known as the {\em 51\% attack}.
	
	\section{Piggyback Mining - Utilizing an Opposing Selfish Pool}
	\label{sec:piggyback}
	We present piggyback mining, a method enabling the 51\% attack with less than half of the computing power. A miner can choose which block to mine on, which transactions to include, and when to publish a successfully mined block. These freedoms allow formal definitions of legitimate and deviant strategies.
	 
	\begin{definition}[Protocol-Legitimate and Deviant Strategies]
    \label{def:deviant}
	A {\em protocol-legitimate strategy} is a strategy that always mines on the longest public branch and immediately publishes blocks it finds. Strategies that are not protocol-legitimate are called {\em deviant strategies}.
	\end{definition}
	
	Examples of protocol-legitimate strategies include the default strategy, and PettyCompliant \cite{transaction}. 
    Deviant strategies like Selfish mining \cite{main}, Stubborn mining \cite{optimal,stubborn}, and others \cite{transaction,unclemaker}, have been proven to offer better rewards than their protocol legitimate counterparts under certain conditions. 
	
	We next define a useful property, named {\em slowdown} and show this property to be characteristic of any deviant strategy.
	
	\begin{definition}[Slowdown]
	A strategy causes a {\em slowdown} if, when a miner or pool implements it, the average progress rate of the main branch is slower than the average progress rate when all miners and pools were protocol-legitimate. 
	\end{definition}
	
	Slowdown can be measured in terms of discarded blocks. Blocks are discarded even if all miners use protocol-legitimate strategies, yet their number increases with a slowdown.
	\begin{samepage}
	\begin{theorem}
	\label{th:slowdown}
    Every deviant strategy causes a slowdown.
    \end{theorem}
    \end{samepage}
    Before proving Theorem~\ref{th:slowdown} (Section~\ref{sec:proof}) we first demonstrate its importance, by demonstrating how a pool can utilize this slowdown to gain complete control of the main branch and commit double spending attacks.
	\subsection{The Piggyback Strategy}
	\label{sec:piggystrat}
	Assume that a pool $P$ detects a second pool $P_s$ that uses a selfish mining strategy~\cite{detect}. Let $P_h$ be the pool of honest miners. and $|P|$ be the computing power of $P$, represented as percentage of the total computational power of all miners in the network. For some values of $|P|$, the slowdown caused by the selfish miners creates a situation where $P$'s rate of progress is faster than the rate of progress of all the other miners, thus giving $P$ an option to gain complete control of the main branch, like the 51\% attack. We call this the \emph{piggyback} strategy. We show next its feasibility by analyzing two complementary cases based 
	 on the values of $|P_s|$ and $|P_h|$. 
\subsubsection{$|P| > |P_s| > |P_h|$}
	$P$ can apply a piggyback strategy by always mining on a private branch, not revealing their blocks.
    After some time, because $P_s$ has more computing power than honest miners, they will gain a lead, ensuring that both $P_s$ and the honest pools always mine on different branches, causing a slowdown. This also ensures that the public branch will eventually only contain blocks mined by $P_s$.
    Finally, since $P$ has more computing power than $P_s$, the branch that $P$ mined secretly will be longer than the public branch (effectively only mined by $P_s$), so $P$ can reveal their branch, which will be accepted as the main branch according to the bitcoin protocol, thus wasting all work done by $P_s$ and $P_h$.
	
	\begin{figure}[t!]
    \includegraphics[width=8cm]{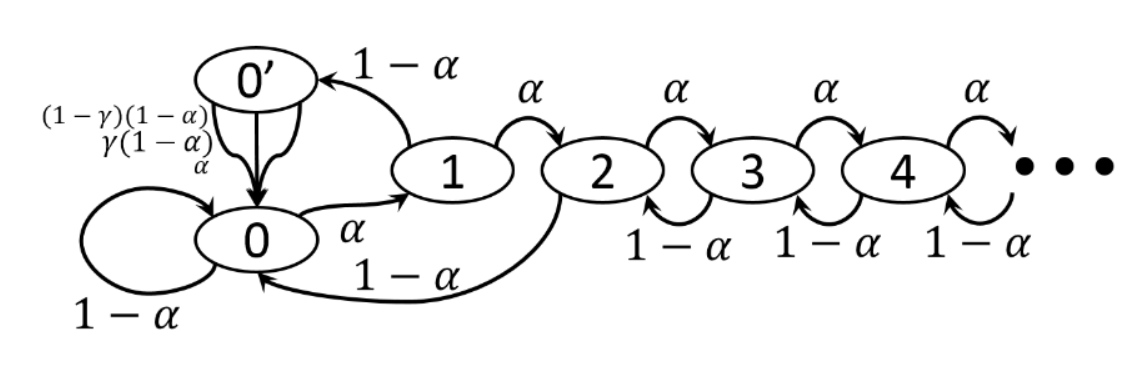}
        \centering
    \caption{The Markov chain as envisioned by Eyal and Sirer \cite{main}. States represent the lead of a secret branch over the public-main branch, $\alpha$ represents the selfish pool's relative mining power, 
    and $\gamma$ represents the portion of honest miners who will mine on top of the selfish branch in case of a competition.}
    \label{state}
    \end{figure}
	\subsubsection{$|P| > |P_h| \ge |P_s|$}
	$P_s$'s selfish mining may not 'take over' the public branch, but it will create a slowdown that can benefit $P$. 
		So long as $P$ continues to withhold their blocks, the chain reverts to the case of a single selfish pool, while the rest are honest~\cite{main}. This means, that in order to calculate the slowdown, we need to normalize $|P_s|$ and $|P_h|$, and then we can proceed to use the calculations made in \cite{main}. A slowdown is caused when a mined block gets discarded, which happens in one of the following three scenarios (see Fig.~\ref{state} for illustration), separated according to the size of the lead of the selfish pool.
	
	\begin{compactitem}
	    \item {\em A lead of one block:} Once the honest miners mine a block, the selfish pool will publish theirs. As a result of the contest in state $0'$, one block will be discarded (belonging to either the honest miners or the selfish pool).
	    \item {\em A lead of two blocks:} Once another honest block is mined, the selfish miners will publish both their blocks, causing the honest one to be discarded.
	    \item {\em A lead of more than two blocks:} Every block mined by the honest miners is discarded when the selfish miners reveal another of their secret blocks. 
	\end{compactitem}
	Eyal and Sirer~\cite[p.09]{main} use a Markov chain (Fig.~\ref{state}) to calculate the probability of being in each state. State $0'$ represents the case where the selfish pool have published their entire secret chain, and are in competition with the main chain. Note that the move from $0'$ to $0$ is with probability 1. The separate transfers represent the different options of getting there, i.e., who wins the contest. However, we are only interested in the progress speed, and not who controls the chain, so the separate arrows are irrelevant for our analysis. The discussion above and the Markov chain lead us to conclude that a block is only discarded when the honest miners mine a block, and the only states in which a block is not discarded are $0$ and $0'$. 
	
	In the piggyback strategy, $P$ withholds all of its blocks until the optimum time to reveal them (discussed in Section \ref{sec:optimal}). So long as $P$ continues to withhold, the system effectively reverts to a system where the total computing power is $|P_s| + |P_h|$. In the reduced system, we have a single selfish pool.
	
	Let $\alpha=\frac{|P_s|}{|P_s| + |P_h|}$ define $P_s$'s relative computing power in the new system. 
	Since we now have a system identical to the one discussed in \cite{main}, we can adopt their calculated probabilities and use them and $\alpha$ to calculate the progress rate.
	The probability of being in a state where a block could be discarded is $(1-p_0-p_{0'})$. We multiply this by the probability that the honest miners mine a new block $(1-\alpha)$, and finally adopt the Markov chain calculations for the probabilities $p_0=\frac{\alpha-2\alpha^2}{\alpha(2\alpha^3-4\alpha^2+1)}$, and $p_{0'}=\frac{(1-\alpha)(\alpha-2\alpha^2)}{2\alpha^3-4\alpha^2+1}$ to get
    \[1-(1-\alpha)(1-p_0-p_{0'}) = 1-\frac{\alpha(1-\alpha)^2}{2\alpha^3-4\alpha^2+1}\]
	This equation captures the progress rate of a system containing only one selfish pool. To get the relative computing power of their branch in our system (containing $P$, $P_s$, and honest) we multiply by their relative size, to get the effective relative computing power of $P_s$ and $P_h$ combined:
	\begin{equation}\label{eq:1}
	    (|P_s|+|P_h|)(1-\frac{\alpha(1-\alpha)^2}{2\alpha^3-4\alpha^2+1})
	\end{equation}
	\begin{corollary}
	For $P$ to guarantee a successful piggyback attack, it must hold that $|P|>(|P_s|+|P_h|)(1-\frac{\alpha(1-\alpha)^2}{2\alpha^3-4\alpha^2+1})$.
    \end{corollary}

    \begin{figure}[!t]
    \centering
    \includegraphics[width=0.43\textwidth]{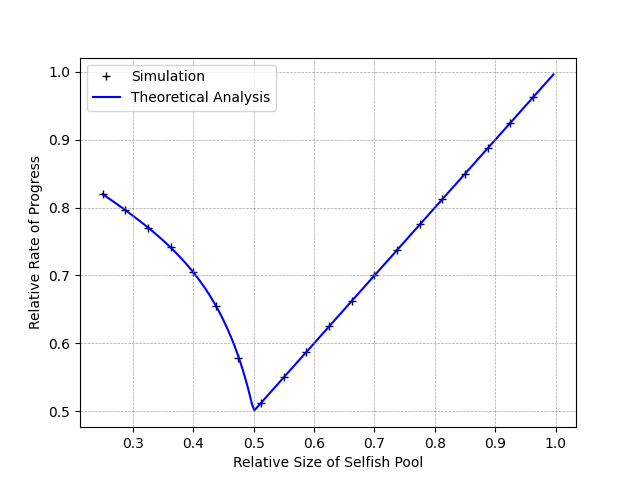}
    \caption{Relative selfish pool size compared to the slowdown caused by it. Simulation and theoretical analysis show that selfish mining always causes a slowdown, where the most significant slowdown is when the selfish pool is the same size as the honest miners.}
    \label{beta}
    \end{figure}

     \begin{figure}[!t]
    \centering
    \includegraphics[width=0.43\textwidth]{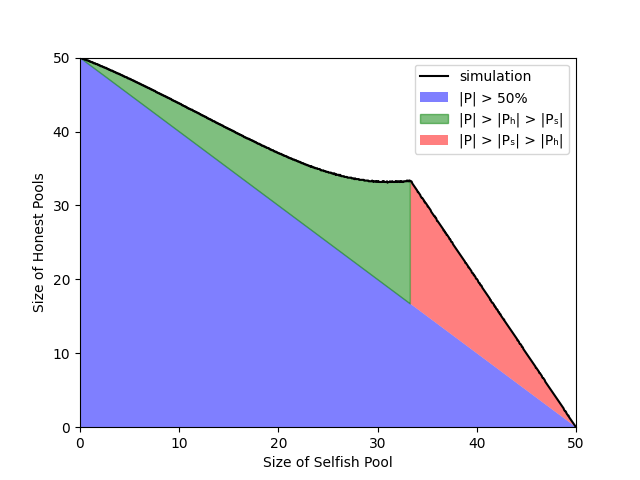}
		\caption{For a given relative size of the selfish pool, the black line shows the maximal size of honest miners for which a piggyback attack is possible. Colored areas are where piggybacking is possible, and each colored area represents one of the cases described in this section. 
        }
    \label{range}
    \end{figure}
    

    To corroborate our mathematical findings, we ran a simulation. We fixed the size of the selfish pool, and then calculated the size of the branch in regards to the amount of blocks mined, to measure the slowdown as a function of $\alpha$. Fig.~\ref{beta} shows that the two graphs align perfectly. We ran an additional simulation to find the ranges of $|P_s|$ and $|P_h|$ where $P$ can effectively take over the blockchain. We fixed the size of the selfish pool, and then found the minimal size of $P$ where it can gain control of the blockchain. In Fig.~\ref{range}, the range below the simulation line is divided into three different areas. In the blue area, $P$ holds more that $50\%$ of the mining power, this is the classical $51\%$ attack, the red area represents the scenario of $|P| > |P_s| > |P_h|$,
    and the green area is where $P_s$ is smaller than the pools of honest miners and $P$ utilizes the slowdown in the chain progression to mount its attack. The black line shows the simulated lower bound on $\alpha$, with which $P$ can pull off the piggyback attack. This line matches the result of the theoretical analysis shown in Eq.~\ref{eq:1}.

    \subsection{Extending Piggyback Mining to Any Deviant Strategy}
    \label{sec:proof}
    We first prove Theorem~\ref{th:slowdown} and then show that for every deviant pool, exists $0<i<0.5$ such that a pool of size $i$ can successfully implement piggybacking.
    
    \begin{proof}
    From Definition~\ref{def:deviant} we know that every deviant strategy either 1) mines on older blocks or 2) hides blocks that it finds.

    The first group of strategies, containing Undercutting \cite{transaction} and Stubborn mining \cite{optimal,stubborn}, mine on older blocks, and when finding a block before honest miners, publishing that block and wasting resources. Denote $B_{-i}$ as the $i^{th}$ block behind the head of the main branch. For strategies in this group there is a probability $\rho>0$ that the pool will mine on block $B_{-j}$ for some $j > 0$, and will succeed and find a block $B_{d}$ on it. In this case, either $B_{-j+1}$ or $B_{d}$ must be discarded, causing a slowdown.

We have already proven that every strategy that mines on older blocks causes a slowdown so we only need to prove the theorem for strategies that always mine on the newest block. From Definition~\ref{def:deviant}, any remaining deviant strategy belongs to the second group (e.g., the Selfish mining strategy~\cite{main}), strategies that do not reveal their mined blocks immediately. For strategies like this there is a probability $\rho >0$ for which the deviant pool mines a block $B_{d}$ and does not publish it until an honest miner publishes a block $B_{h}$ of their own. In this case, either $B_{d}$ or $B_{h}$ must be discarded, consequently causing a slowdown. 
\end{proof}

    Section~\ref{sec:piggystrat} shows how the piggyback strategy works against selfish mining, but the only feature of selfish mining we used
    is the slowdown selfish mining causes. Thus, the theorem gives rise to the following generalizing corollary.
    
    \begin{corollary}
    The piggyback strategy works against every deviant strategy.
    \end{corollary}
    
    \section{Optimal Waiting Time for Piggybackers to Reveal Their Branch}
    \label{sec:optimal}

    
Using piggybacking against a selfish pool immediately boosts the its profits, unlike selfish mining where gains are delayed~\cite{main}. A selfish mining pool creates forks to gain more profit, effectively diverting rewards away from other miners, including the piggybacking group. However, piggybacking prevents the selfish pool from successfully forking against the piggybacking's branch, quickly restoring lost revenue and increasing profits. While the piggybacking strategy ensures that, eventually, the piggybacker's branch is longer than all other branches, predicting the exact moment of occurrence is challenging.

 Piggybacking pools have an incentive to reveal their branch as soon as possible, since delays could allow the selfish mining pool to notice their strategy and switch to a protocol-legitimate strategy.    
However, if the piggybacker's branch is revealed too soon, there is a risk that the current main branch (or the secret branch in selfish mining~\cite{main}) could still be longer. In that case, the piggybackers will not be able to capture the full profit.
    
    We analyze the probability that the piggybacker's branch is the longest, based on their relative computing power ($\beta$) and the total number of mined blocks (M). Since many blockchain systems adjust their difficulty so that a new block is mined at regular intervals, this gives a sound approximation to the wait time 
    before revealing a branch.

    A block has a $\beta$ chance to be mined by the piggybacking pool, and consecutive attempts are independent, so we can treat the amount of blocks mined by the piggybacking group as a binomial random variable with probability $\beta$, yielding
    \begin{equation}
    \mathbb{P}[\mbox{longest branch}]=\mathbb{P}[\operatorname{Bin}(\beta,M) > \frac{M}2]
    \end{equation}
    
    

    
    \begin{figure}[htpb]
        \centering
            \includegraphics[width=0.43\textwidth]{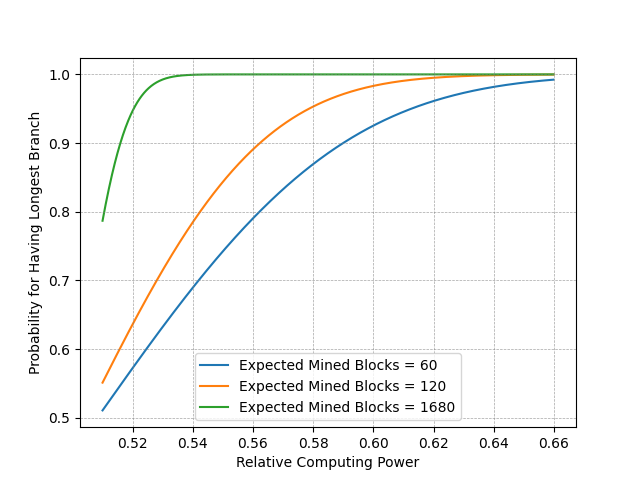}
            \caption{The probability of maintaining the longest blockchain branch as a function of the relative computing power of a piggybacking pool, given a fixed number of total mined blocks.}
            \label{time}
    \end{figure}
    
    Figure~\ref{time} shows the dependency between $\beta$, 
    expected $M$ (mined blocks), and $\mathbb{P}[\mbox{longest branch}]$. 
    For almost every $\beta$ value, 
    the chance of having the longest branch after 1680 blocks have been mined (approximately two weeks in bitcoin's case) is extremely high (note that the y-axis in Figure~\ref{time} starts at 50\%). Even after only 120 mined blocks (20 hours in bitcoin's case) the risk is low for most relative mining speeds.
    
    \section{Responding to a Piggybacking Strategy}
    \label{sec:resilience}
    The coexistence of a deviant pool and an active piggybacking pool is not a state of equilibrium, as the deviant pool can increase its own profit by becoming honest and thereby invalidating the piggybacking pool. As a result, the presence of a piggybacking pool acts as a deterrent to deviant mining, decreasing their revenue.
    This shift benefits honest miners by increasing their overall share of the rewards. In Section~\ref{sec:piggyback} we introduced the notion of a slowdown. We say that a slowdown is a {\em sufficient slowdown} if applying a piggybacking strategy once such a slowdown is detected yields an outcome where all profit goes to the piggybacking pool.
    
    
    \begin{definition}[Resilience]
    Let $P_s$ and $P_o$ be pools, implementing strategies $s$ and piggybacking, respectively.
    The {\em resilience} of a pool $P_s$ (denoted $Res(|P_s|,s)$), is the smallest size $|P_o|$ that can successfully implement the piggybacking strategy.
	\end{definition}
    \begin{sloppypar}
    For example, the resilience of honest mining is always $0.5$ and $Res(0.4, SELFISH) = 0.4$ as was calculated in Section~\ref{sec:piggystrat}, where $SELFISH$ is selfish mining~\cite{main}.
\end{sloppypar}

    Denote the space of all strategies as $S$. Denote the deviant pool as $P_d$ and denote $RelRev(|P_d|,s)$ as the relative revenue of $P_d$ when implementing $s\in S$. Finally, we denote the largest 
    Piggybacking pool as $P_o$. Let $Str$ be the set of all strategies that $P_d$ can implement without triggering $P_o$, formally,
\[Str(|P_d|,|P_o|) = \{s\in S| Res(|P_d|,s) > |P_o|\}.\] The maximum revenue $P_d$ can get is \[MaxRev(|P_d|, |P_o|) = \sup\{RelRev(|P_d|,s)| s\in Str(|P_d|,|P_o|)\}.\] 

The following corollary offers the desired property that a sufficiently large honest pool keeps other pools from implementing most deviant strategies. 

      \begin{corollary}\label{th:piggybacking}
        The existence of a
        piggybacking pool reduces the relative revenue of deviant pools.
    \end{corollary}

Corollary~\ref{th:piggybacking} follows immediately from the observation that $MaxRev(P_d, P_o)$ is inversely monotone with respect to $|P_o|$ and so $MaxRev(P_d, P_o) \le MaxRev(P_d, 0)$. 
 
    \section{Conclusions}
    \label{sec:conclusions}

    We propose the {\em piggyback strategy}, which exploits the slowdown inherent in deviant (malicious) strategies to collect rewards. We have proven that piggybacking can counter any deviant strategy and, consequently, disincentivizes other pools from adopting such strategies.
    This paper offers a theoretically grounded basis to understand why selfish mining has not yet been seen in blockchain ecosystems in the wild. 
    
	\bibliographystyle{plain}
    \bibliography{main}
\end{document}